\documentstyle[aaspptwo,psfig]{article}
\def\etal{{\it et al. }}

\begin{document}

\title{Globular Clusters in the Sombrero Galaxy (NGC 4594)
\altaffilmark{1,2}}

\author{Duncan A. Forbes}
\affil{Lick Observatory, University of California, Santa Cruz, CA 95064}
\affil{and}
\affil{School of Physics and Space Research, University of Birmingham,
Edgbaston, Birmingham B15 2TT, United Kingdom}
\affil{Electronic mail: forbes@lick.ucsc.edu}

\author{Carl J. Grillmair}
\affil{Jet Propulsion Laboratory, 4800 Oak Grove Drive, Pasadena, CA 91109}
\affil{Electronic mail: carl@grandpa.jpl.nasa.gov}

\author{R. Chris Smith}
\affil{Department of Astronomy, University of Michigan, 934 Dennison
Building, Ann Arbor, MI 48109}
\affil{Electronic mail: chris@astro.lsa.umich.edu}

\altaffiltext{1}{Based on observations obtained at Cerro Tololo
Inter--American Observatory. CTIO is operated by AURA, Inc. under
contract to the National Science Foundation.} 
\altaffiltext{2}{Lick Bulletin 1350}

\begin{abstract}

The Sombrero galaxy, NGC 4594, contains the most numerous globular
cluster system of any nearby spiral. It is an ideal candidate in which
to study the globular clusters and contrast them with those in Local
Group spirals. Here we present B and I imaging from the CTIO Schmidt
telescope which gives a field-of-view of 31$^{'}$ $\times$ 31$^{'}$. 
Using DAOPHOT
we have detected over 400 globular clusters and derived their
magnitudes, B--I colors and photometric metallicities. We have
attempted to separate our
sample into disk and bulge/halo globular cluster 
populations, based on location in the galaxy. 
There is some evidence that the disk population is more
metal--rich than the bulge/halo globular clusters, however
contamination, dust reddening and small number statistics makes this
result very tentative. 
We find that the median metallicity of the bulge/halo
globular clusters is [Fe/H] = --0.8. 
This metallicity is consistent with previous estimates based on 
smaller samples. It is also similar to the metallicity 
predicted by the globular cluster metallicity -- galaxy luminosity 
relation. 
As with our Galaxy, there is no radial metallicity gradient in
the halo globular clusters. This suggests that the spheriodal
component of NGC 4594 did not form by a dissipational process. 

\end{abstract}

\section{Introduction}

Measurement of the abundances of Milky Way globular clusters (GCs)
have provided important clues on the formation and chemical enrichment
history of the Galaxy. Combined with kinematics, a clear dichotomy of
metal--rich disk GCs and metal--poor halo GCs have been identified
(e.g. Zinn 1985). 
The disk and halo GCs have a mean metallicity of 
[Fe/H] = --0.6 and --1.6 respectively (e.g. Ashman \& Bird 1993). 
Globular clusters in our sister galaxy, M31, have
been relatively well--studied so that spectroscopic metallicity 
measurements are available for about half of them (Huchra, Brodie \&
Kent 1991). 
A bimodal metallicity distribution is not as obvious for M31 but the
analysis of Ashman \& Bird (1993) indicates peaks at 
[Fe/H] = --0.6 and --1.5, i.e. similar to the Milky Way. 
The other Local Group spiral, M33, has about
30 known GCs, of which spectroscopic metallicities have been obtained
for 22 (Brodie \& Huchra 1991). The mean metallicity of [Fe/H] =
--1.55 $\pm$ 0.37 is similar to that of Milky Way and M31 halo
GCs. The small number of GCs makes it difficult to
convincingly identify a population of disk GCs. 

Beyond the Local Group the nearest spirals are M81 (m--M = 27.8)
and NGC 4594 (m--M = 29.73). Perelmuter \& Racine (1995) estimated from
imaging of M81 that it contains $\sim$ 200 GCs. 
Spectra for 30 GCs
have been obtained (Brodie \& Huchra 1991; Perelmuter, Brodie \&
Huchra 1995) which give a mean [Fe/H] = --1.48 $\pm$ 0.19.
The Sombrero galaxy, NGC 4594, is a giant Sa galaxy (M$_V$ = --22.0)
with almost 2,000 GCs (Bridges \& Hanes 1992). This is significantly
more than either the Milky Way with N$_{GC}$ = 160 $\pm$ 20 or M31
with N$_{GC}$ = 350 $\pm$ 100 (Harris 1996).
It is also nearly edge--on ($i = 84^{\circ}$) so that 
GCs projected close to the galaxy major axis
can be used to define a plausible sample of disk GCs. Separating disk
and halo GCs in M31 ($i = 38^{\circ}$) has been a major source of
uncertainty. Thus NGC 4594 is an ideal candidate in 
which to study the GC system of another spiral galaxy.  

The distance to NGC 4594, from the average of the surface brightness
fluctuation and planetary nebulae luminosity function methods, 
is 8.8 $\pm$ 0.4 Mpc (Ciardullo, Jacoby \& Tonry 1993). At this distance
GCs range in magnitude from B $\sim$ 18--27, so that only the
brightest few GCs are accessible spectroscopically by today's large
telescopes. A valiant effort was made recently by Bridges \etal
(1996). After 7 hours of integration time over 
4 nights of observing with the 4.2m
William Herschel Telescope, they obtained spectra of 34 confirmed
GCs. Unfortunately their spectra were not of sufficient quality to
determine a metallicity for individual GCs. Summed together they
derived a mean of [Fe/H] = --0.7 $\pm$ 0.3.
An alternative approach, is to obtain GC colors from imaging studies.
Imaging has the advantage of
being very efficient and photometric metallicities can be derived from
the Galactic GC color--metallicity relation (Couture \etal 1990). 

The GC system of NGC 4594 has been studied photometrically by 
Harris \etal (1984) and Bridges \& Hanes (1992). The first of these
was a wide field-of-view photographic study in U and V bands. They
examined the spatial distribution and estimated the GC specific
frequency. The second study targeted three areas of the
galaxy and imaged these in B and V bands with a 2.2$^{'}$ $\times$ 3.6$^{'}$ 
CCD. From the B--V colors of 131 GCs they derived a 
mean [Fe/H] = --0.81 $\pm$ 0.25
for the GC system. 
Here we present 31$^{'}$ $\times$ 31$^{'}$ field-of-view CCD imaging of
NGC 4594 in B and I bands. We measure B--I colors which are about
twice as sensitive to metallicity as B--V colors (e.g. Geisler, Lee \&
Kim 1996). Our wide field-of-view imaging allows us to define
spatially samples of `disk' and `bulge/halo' GCs, and examine their 
metallicity distributions.

\section{The Data and Globular Cluster Selection}

Broad--band B and I images of NGC 4594 were taken with the 
Cerro Tololo Interamerican Observatory (CTIO) Schmidt telescope. 
We used a Thomson 1024 $\times$ 1024 array with a pixel
scale of 1.84$^{''}$/pixel. The images were obtained in 1994 February
12 and 13. 
Reduction was carried out in the standard
way (i.e. bias and
dark subtraction, flat--fielding and sky subtraction). 
The total exposure times were 7200 and 2400 secs 
for the B and I images respectively. 
After combining, the B image was 
calibrated using aperture photometry from the UBV catalog of 
of Longo \& de Vaucouleurs (1983) which gave an rms accuracy of $\pm$
0.06 mag. Burkhead (1986) found that the bulge had a near constant
color of B--I = 2.7 $\pm$ 0.1 and this was used to calibrate the I
image. Finally, the magnitudes are corrected for Galactic extinction
of A$_B$ = 0.12 and A$_I$ = 0.05 (Bender \etal 1992). 
Typical photometric errors are 0.08 at B = 20.5 and 0.18 at B = 21.5. 
Our B--I colors are transformed into [Fe/H] values using the
color--metallicity relation of Couture \etal (1990) which is based on
Milky Way GCs. 
This relation is calibrated from about 1/100 solar metallicity up to
solar. 
The typical random error in this transformation 
is $\pm$ 0.15 dex in [Fe/H]. 
From the final images, we subtracted a smooth elliptical model for the
bulge using IRAF software. 

Contamination from foreground stars and background galaxies is a
concern for most GC imaging studies. Although not ideal, one method is
to obtain similar exposure time images in some nearby `blank' field
and process these images in the same way as the actual data. This
provides a statistical contamination correction. We do not have such a 
`blank' field, so we attempt to estimate the contamination using
star and galaxy surface densities from the literature. 
Before applying this method we will restrict candidate GCs to
magnitudes of  17.5 $<$ B $<$ 22.
The faint limit is a little  
brighter than the expected
turnover of the GC luminosity function (i.e. B $\sim$ 22.5) and was
chosen to ensure that a color bias was not introduced.  
The bright limit is at least 3$\sigma$ from the turnover
magnitude assuming a dispersion characteristic of the Milky Way and
M31 (Secker \& Harris 1993), while excluding the most obvious foreground
stars. 

The number of foreground stars can be
estimated from the Bahcall \& Soneria (1981) model of the Milky Way. 
For our 31$^{'}$ $\times$ 31$^{'}$ image in the direction of NGC 4594,
their model predicts about 750 stars with 17.5 $<$ B $<$ 22. 
Using figure 1 of Koo
\& Kron (1992), we estimate our image contains about 220 background
galaxies down to B = 22. Thus the total contamination in our image
from stars and galaxies, before any color selection, is about 970 for
magnitiudes 17.5 $<$ B $<$ 22. To compare this number with the number
of objects in the B image, we have used DAOPHOT with a 
fairly conservative detection threshold of 
5$\sigma$ per pixel. This gives $\sim$ 1400 objects with 
17.5 $<$ B $<$ 22. Thus we estimate our image contains $\sim$ 430 {\it
bona fide} GCs. 
Based on our detection limit and 
an expected turnover magnitude of B $\sim$ 22.5, our sample represents
about 1/3 of the total GC population. So the total number of GCs is
$\sim$ 1300, which is at the lower limit of the range estimated by 
Bridges \& Hanes (1992) i.e. 1900 $\pm$ 600. 
We note that although we have only detected a fraction of the total GC
population, there is no known correlation of GC metallicity with luminosity
or mass (Huchra, Brodie \& Kent 1991) which might otherwise have bias
our results. 

Although stars and galaxies cover a similar range of B--I colors as
GCs, we know from other studies that the vast majority of GCs 
have metallicities --2 $<$ [Fe/H] $<$ +0.5 (i.e. 0.15 $<$ B--I $<$
2.33). So excluding objects
beyond these values  will preferentially 
remove stars and galaxies from the object lists. After applying this
metallicity (color) selection and making restrictions 
in the DAOPHOT sharpness and roundness parameters, our sample consists
of 457 candidate GCs. This is similar to the $\sim$ 430 objects left
after a statistical correction for stars and galaxies has been
applied. From these arguments we expect that the majority of objects 
in our final list are indeed GCs.

Next we attempt to divide the sample into `disk' and 
non--disk (which we will call `bulge/halo') GCs. 
In our Galaxy (Gilmore \& Reid 1983) and NGC 4594 (Burkhead 1986) the
scale height of disk stars is Z $\sim$ 1 kpc. Disk GCs in our Galaxy
can reach distances above the plane of Z $\le$ 4 kpc. We have chosen to
define disk GCs in NGC 4594 as those with scale heights 
$\vert$Z$\vert$ $\le$ 4 kpc. 
The colors of the disk objects may be reddened by the galaxy's dust
lane. To minimize this effect we have 
excluded objects that are close to the central dust i.e., objects with
major axis distances of less than 
5 kpc of the galaxy center. 
Conversely bulge/halo GCs are simply 
defined as those with $\vert$Z$\vert$ $>$ 4 kpc; it is not clear
whether such GCs are associated with the dominant bulge or the halo. 
Dividing the
sample in this way we have 79 disk objects and 378 halo objects. 
Due to projection effects
some bugle/halo GCs will be present in the disk sample. From the surface
density of halo GCs we estimate that between 1/3 and 1/2 of the disk
GCs are not associated with the disk.

\section{Results and Discussion}

In this section, we first 
discuss the GC sample of Bridges
\etal (1996). Second, 
we examine the GC metallicity distribution in NGC 4594 and finally 
we search for spatial abundance gradients in this distribution.  

After superposing the position of the Bridges \etal  
34 spectroscopically--confirmed GCs on our
CCD images we managed to measure B and I magnitudes
for 24 (the others are presumably fainter than our limiting
magnitude). The mean color is B--I = 1.9 $\pm$ 0.2 
(error on the mean), which corresponds
to [Fe/H] = --0.7 $\pm$ 0.5. 
This is reassuringly consistent with the
spectroscopic mean metallicity, for all 34 GCs, of [Fe/H] = --0.7
$\pm$ 0.3. This suggests that 
the transformation from
color to metallicity is reasonable.

In Fig. 1 we show the metallicity distribution for disk 
($\vert$Z$\vert$ $\le$ 4 kpc) and bulge/halo 
($\vert$Z$\vert$ $>$ 4 kpc) objects. 
The median metallicity of the bulge sample is [Fe/H] =
--0.8. 
This is similar to that obtained by 
Bridges \etal (1996) and 
suggests that our sample is
dominated by {\it bona fide} GCs. 
As pointed out by Bridges \etal the GC metallicity is more comparable to
that found in giant ellipticals than other spirals. However this might 
be expected given the claimed correlation of GC metallicity with parent galaxy
luminosity (e.g. Brodie \& Huchra 1991). Using the recent best--fit of 
Forbes \etal (1996) and the bulge luminosity of NGC 4594, the relation
predicts a GC system mean metallicity of [Fe/H] = --0.6
$\pm$ 0.3 (rms error). 
The disk sample has a median metallicity of [Fe/H] = --0.5. As the
disk sample will include a significant fraction of bulge GCs, the
average metallicity of the disk sample is likely to be somewhat higher
than --0.5. 
If we separate the 24 Bridges \etal (1996) GCs, with
available B--I colors, into bulge and disk systems, then 
their metallicities are [Fe/H] =  --1.2 
and --0.1 respectively. Thus there is some tentative evidence, from 
both our sample and that of Bridges \etal that NGC 4594 contains a 
population of
disk GCs that are slightly more metal--rich than non--disk GCs. 
Although it is difficult to rule out preferential reddening of the GCs
by dust, we do not think this has had a strong effect on the disk GC
colors. First, we have excluded the inner galaxy regions with the
obvious dust lane. Second, we do not find a strong radial GC color
gradient in either scale height (see Fig. 2) or along the major axis as
might be expected from centrally concentrated dust. 
The possibility that inner GCs are instrisically blue (and so removing
a reddening trend due to dust) is very unlikely given studies of other
galaxies (e.g. Harris 1991; Geisler \etal 1996). 

Abundance gradients, or lack of, provide useful constraints to galaxy
formation models. For example, the absence of an abundance gradient in
Milky Way halo GCs supports the view of Searle \& Zinn (1978) in which
the halo is made up of small protogalactic fragments, rather than
arising from a pure dissipational collapse. 
In the M31 GCs the situation is more uncertain. 
Huchra, Brodie \& Kent (1991) found that there exists a upper envelope
in GC metallicity with galactocentric radius but also pointed out that selection
effects and contamination by disk GCs made evidence for an abundance
gradient inconclusive. 
In Fig. 2 we show the spatial variation of GC metallicity for the disk
and bulge/halo samples. 
In the lower panel of Fig. 2 we show GC metallicity 
versus projected galactocentric radius for the bulge/halo objects.
As with the Milky Way GC system, there is no statistically significant
metallicity gradient. This constrains the role of dissipation in
forming the spheroidal component of NGC 4594. 
In a dissipational collapse one expects a metallicity gradient in the
direction of the collapse e.g., perpendicular to a disk. Accretion of
material may reduce the strength of such gradients. 
Our Galaxy
shows a weak trend of GC metallicity with scale height, where the
slope is --0.084 $\pm$ 0.031 dex kpc$^{-1}$ (Armandroff 1993). 
In the upper panel of Fig. 2 we show the metallicity of disk GCs
versus scale height
from the major axis. The data show 
a hint of a weak trend with scale
height, but it is not formally significant.

\section{Conclusions}

After separating a sample of $\sim$ 450 
globular clusters in NGC 4594 with photometric metallicities into
spatially--defined `disk' and `bulge/halo' subsystems, 
we find tentative evidence for a difference in the metallicity of the two
populations.  
The bulge/halo globular clusters have a median metallicity of [Fe/H]
= --0.8 whereas the disk globulars have [Fe/H] $\ge$ --0.5. 
Contamination, dust reddening and a small number of disk
globular clusters make it difficult to confidently ascribe a higher
average metallicity to the disk population. 
The metallicity of the bulge/halo system is greater than that of the
Milky Way ([Fe/H] = --1.6) or M31 ([Fe/H] = --1.5) halo globulars, 
however it is
consistent with that expected from the correlation of globular cluster
metallicity with galaxy luminosity. 
Like the Milky Way, 
we find no evidence for a metallicity gradient in the bulge/halo
globular cluster population. 
This argues against formation of the spheroidal component by
dissipational collapse.

\noindent
{\bf Acknowledgments}\\
We thank J. Brodie and M. Rabban for their help and useful discussions.  
We also thank the referee, Keith Ashman, for several helpful suggestions.
This research was funded by the HST grant GO-05990.01-94A\\

\newpage
\noindent{\bf References}

\noindent
Armandroff, T. 1993, in The Globular Cluster--Galaxy Connection,
edited by G. Smith and J. Brodie (Astronomical Society of the Pacific,
San Francisco), p. 48\\
Ashman, K. M., \& Bird, C. M. 1993, AJ, 106, 2281\\
Bahcall, J. N., \& Soneira, R. M. 1981, ApJS, 47, 357\\
Bender, R., Burstein, D., \& Faber, S. M. 1990, ApJ, 399, 462\\
Bridges, T. J., \& Hanes, D. A. 1992, AJ, 103, 800\\
Bridges, T. J., Ashman, K. M., Zepf, S. E., Carter, D., Hanes, D. A.,
Sharples, R. M., \& Kavelaars, J. J. 1996, MNRAS, in press\\  
Brodie, J., \& Huchra, J. 1991, ApJ, 379, 157\\
Burkhead, M.S. 1986, AJ, 91, 777\\  
Ciardullo, R., Jacoby, G. H., \& Tonry, J. L. 1993, ApJ, 419, 479\\
Couture, J., Harris, W. E., \& Allwright, J. W. B. 1990, ApJS, 73,
671\\
Forbes, D. A., Franx, M., Illingworth, G. D., \& Carollo, C. M. 1996,
ApJ, 467, 126\\
Geisler, D., Lee, M. G., \& Kim, E. 1996, AJ, in press\\
Gilmore, G., \& Reid, N. 1983, MNRAS, 202, 1025\\
Harris, W. E. 1996, AJ, 112, 1487\\
Harris, W. E., Harris, H. C., \& Harris, G. L. H. 1984, AJ, 89, 216\\
Huchra, J., Brodie, J., \& Kent, S. 1991, ApJ, 370, 495\\
Huchra, J., \etal 1997, in preparation\\
Koo, D. C., \& Kron, R. G. 1992, ARAA, 30, 613\\
Perelmuter, J. M., \& Racine, R. 1995, AJ, 109, 1055\\
Perelmuter, J. M., Brodie, J. P., \& Huchra, J. 1995, AJ, 110, 620\\
Longo, G., \& de Vaucouleurs, A. 1983, A General Catalogue of
Photoelectric Magnitudes in the UBV system, (University of Texas
Press, Austin)\\
Searle, L., \& Zinn, R. 1978, ApJ, 225, 357\\ 
Secker, J., \& Harris, W. E. 1993, AJ, 105, 1358\\
Zinn, R. 1985, ApJ, 293, 424\\


\begin{figure*}[p]
\centerline{\psfig{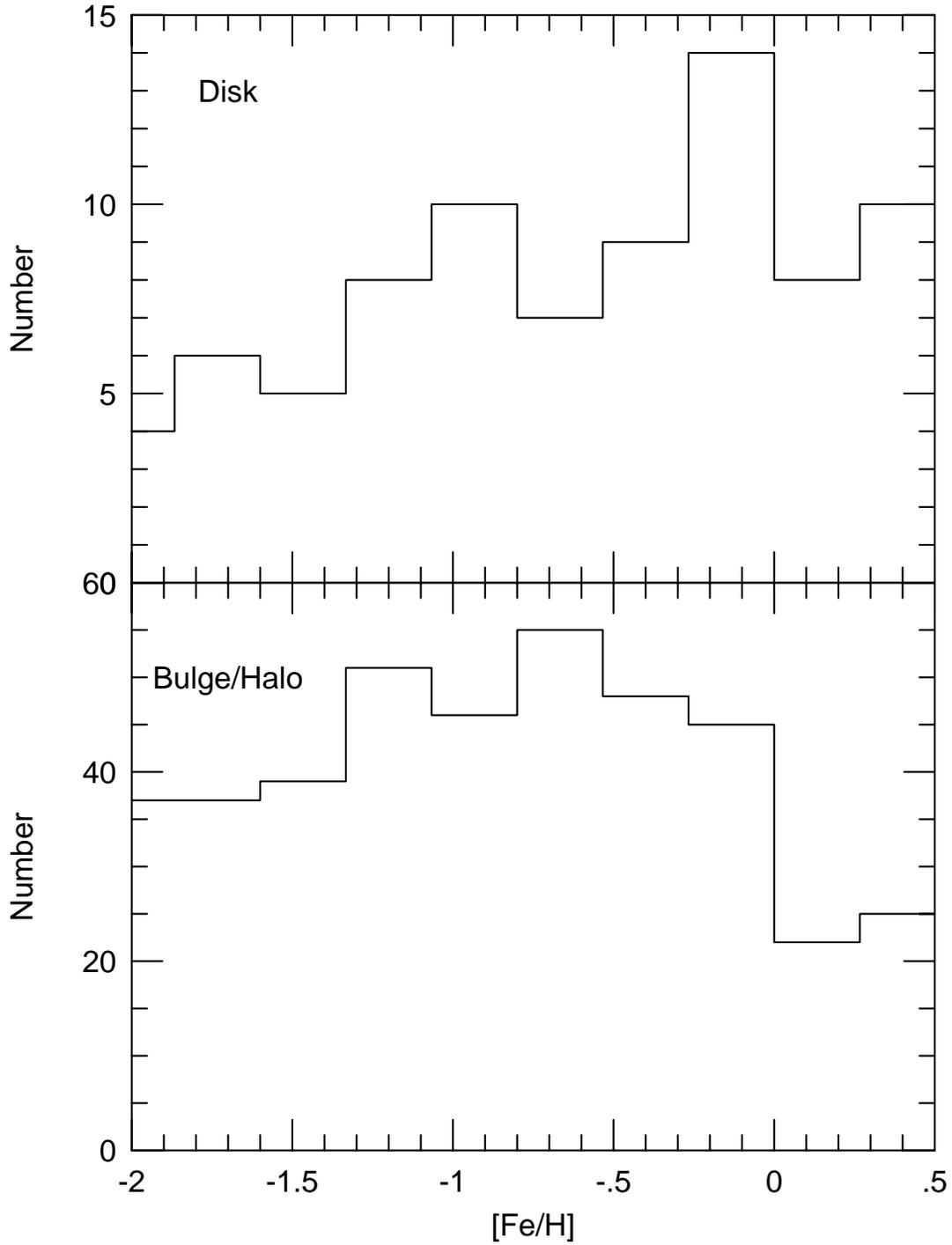}}
\caption{\label{fig1}
Metallicity distribution of globular cluster candidates in NGC 4594. 
The upper panel (the `disk' sample) 
shows objects with scale height $\vert$Z$\vert$ $\le$ 4 kpc 
(excluding the inner galaxy regions). 
The median metallicity is [Fe/H] = --0.5 
although there is clearly a large range in metallicities. 
The lower panel (the `bulge/halo'
sample) shows objects
beyond 4 kpc from the galaxy disk. The median metalliticity is 
[Fe/H] = --0.8. There is
tentative evidence that the disk globular clusters are more metal--rich than the
bulge/halo globular clusters. 
}
\end{figure*}

\begin{figure*}[p]
\centerline{\psfig{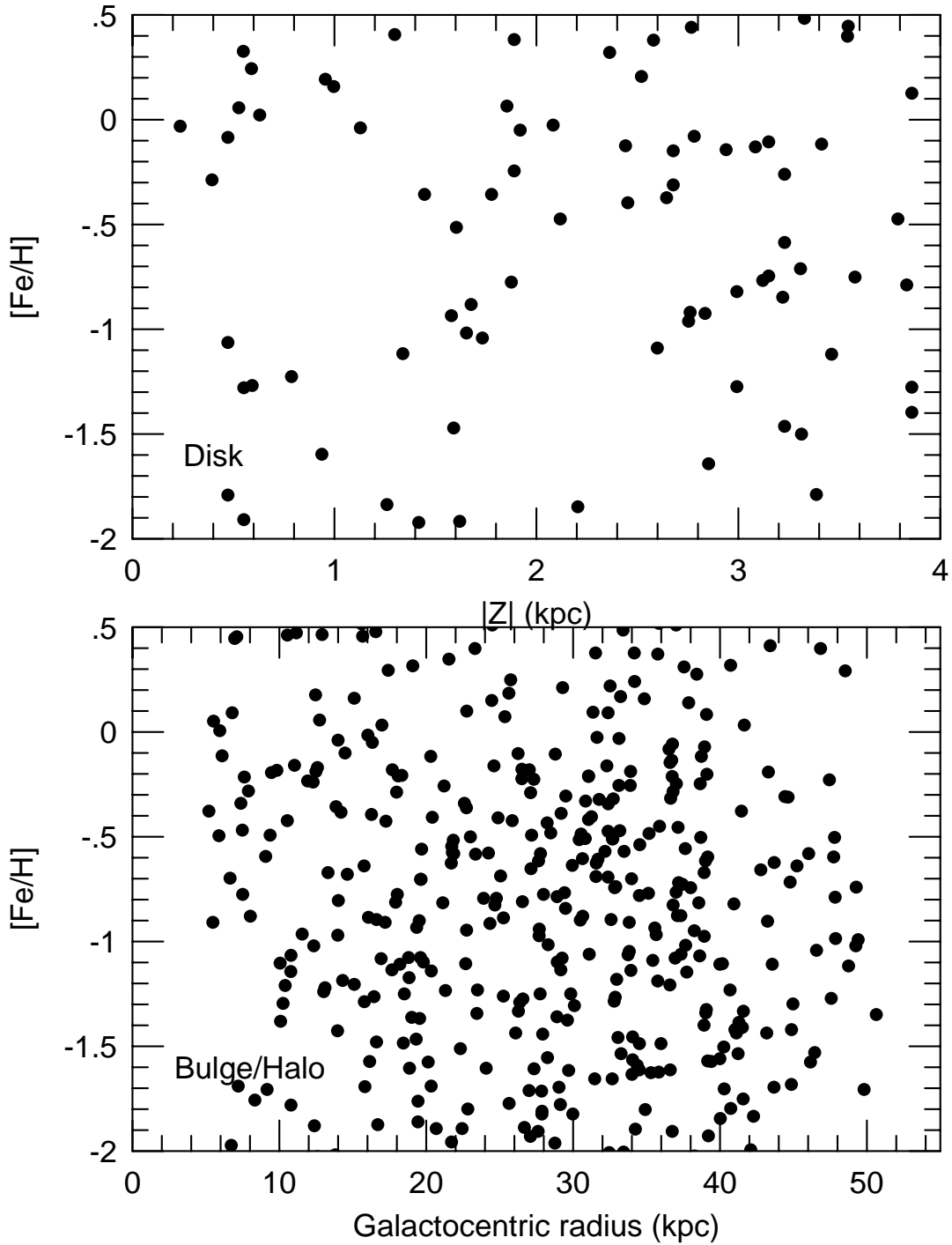}}
\caption{\label{fig2}
Spatial variation of globular cluster metallicity in NGC 4594. 
The upper panel shows the variation of metallicity with scale height for the disk
sample (i.e $\mid$Z$\mid$ $\le$ 4 kpc). There is no statistically
significant gradient.  
The lower panel shows the radial variation of metallicity for the
bulge/halo sample (i.e $\mid$Z$\mid$ $>$ 4 kpc). 
As with the Milky Way, there is no
statistically significant radial
metallicity gradient.
}
\end{figure*}

\end{document}